\documentclass[10pt,conference]{IEEEtran}
\usepackage{courier}
\usepackage[ruled,vlined]{algorithm2e}
\usepackage{graphicx}
\usepackage{paralist}
\usepackage{tabularx}
\usepackage{balance}
\usepackage{multirow}
\usepackage{multicol}
\usepackage{pbox}
\usepackage{mathtools}
\usepackage[TABBOTCAP]{subfigure}
\usepackage{algorithmic}
\usepackage{paralist}
\usepackage{tabularx}
\usepackage{url}
\usepackage{balance}
\usepackage{multirow}
\usepackage{multicol}
\usepackage{amsmath}
\usepackage{setspace}
\usepackage{enumitem}

\usepackage{verbatim}
\usepackage{float}
\usepackage[normalem]{ulem}
\usepackage{xspace}
\usepackage{amssymb}
\usepackage{ifthen}
\usepackage{pifont}
\usepackage{tikz}
\newcommand*\circled[1]{\tikz[baseline=(char.base)]{
            \node[shape=circle,draw,inner sep=0.5pt] (char) {#1};}}
\usepackage{hyperref}
\usepackage[most]{tcolorbox}
\usepackage{booktabs}

\definecolor{MidnightBlue}{HTML}{01693F}
\include{macro}

\newcommand{\ie}{\textit{i.e.,}\xspace}
\newcommand{\eg}{\textit{e.g.,}\xspace}
\newcommand{\etc}{\textit{etc.}\xspace}
\newcommand{\aka}{\textit{aka}\xspace}
\newcommand{\etal}{\textit{et al.}\xspace}

\newcolumntype{C}[1]{>{\centering\arraybackslash}p{#1}}

\begin{document}
%
\title{
It Takes Two to \tango:
Combining Visual and Textual Information for Detecting Duplicate Video-Based Bug Reports
} 


\author{\IEEEauthorblockN{Nathan Cooper\IEEEauthorrefmark{1}, Carlos Bernal-C\'ardenas\IEEEauthorrefmark{1}, Oscar Chaparro\IEEEauthorrefmark{1}, Kevin Moran\IEEEauthorrefmark{2}, Denys Poshyvanyk\IEEEauthorrefmark{1}}
\IEEEauthorblockA{
\IEEEauthorrefmark{1}College of William \& Mary (Williamsburg, VA, USA), \IEEEauthorrefmark{2}George Mason University (Fairfax, VA, USA)\\
nacooper01@email.wm.edu, cebernal@cs.wm.edu, oscarch@wm.edu, kpmoran@gmu.edu, denys@cs.wm.edu
}
}
\maketitle

\begin{abstract}
When a bug manifests in a user-facing application, it is likely to be exposed through the graphical user interface (GUI). Given the importance of visual information to the process of identifying and understanding such bugs, users are increasingly making use of screenshots and screen-recordings as a means to report issues to developers. However, when such information is reported \textit{en masse}, such as during crowd-sourced testing, managing these artifacts can be a time-consuming process. As the reporting of screen-recordings in particular becomes more popular, developers are likely to face challenges related to manually identifying videos that depict duplicate bugs.
Due to their graphical nature, screen-recordings present challenges for automated analysis that preclude the use of current duplicate bug report detection techniques.  To overcome these challenges and aid developers in this task, this paper presents \tango, a duplicate detection technique that operates purely on video-based bug reports by leveraging \textit{both} visual and textual information.  \tango combines tailored computer vision techniques, optical character recognition, and text retrieval. We evaluated multiple configurations of \tango in a comprehensive empirical evaluation on 4,860 duplicate detection tasks that involved a total of 180 screen-recordings from six Android apps. Additionally, we conducted a user study investigating the effort required for developers to manually detect duplicate video-based bug reports and compared this to the effort required to use \tango. The results reveal that \tangos optimal configuration is highly effective at detecting duplicate video-based bug reports, accurately ranking target duplicate videos in the top-2 returned results in 83\% of the tasks. Additionally, our user study shows that, on average, \tango can reduce developer effort by over 60\%, illustrating its practicality.
\end{abstract}


\section{Introduction}\label{sec:introduction}

Many modern mobile applications (apps) allow users to report bugs in a graphical form, given the GUI-based nature of mobile apps. For instance, Android and iOS apps can include built-in screen-recording capabilities in order to simplify the reporting of bugs by end-users and crowd-testers~\cite{bugclipper,testfairy,bugsee}. The reporting of visual data is also supported by many crowd-testing and bug reporting services for mobile apps~\cite{bugclipper,testfairy,ubertesters,Instabug,bugsee,snaffu,bugreplay,bugsquasher,birdeatsbugs,outklip}, which intend to aid developers in collecting, processing, and understanding the reported bugs~\cite{3Bettenburg:FSE08,Mao:ASE17}.

The proliferation of sharing images to convey additional context for understanding bugs, \eg in Stack Overflow Q\&As, has been steadily increasing over the last few years \cite{Nayebi2020}. Given this and the increased integration of screen capture technology into mobile apps, developers are likely to face a growing set of challenges related to processing and managing app screen-recordings in order to triage and resolve bugs --- and hence maintain the quality of their apps. 

One important challenge that developers will likely face in relation to video-related artifacts is determining whether two videos depict and report the same bug (\ie detecting duplicate video-based bug reports), as it is currently done for textual bug reports~\cite{Rakha:TSE'18,Rakha2018,32Bettenburg:ICSM08}. When video-based bug reports are collected at scale, either via a crowdsourced testing  service~\cite{bugclipper,testfairy,ubertesters,Instabug,bugsee,snaffu,bugreplay,bugsquasher,birdeatsbugs,outklip} or by popular apps, the sizable corpus of collected reports will likely lead to significant developer effort dedicated to determining if a new bug report depicts a previously-reported fault, which is necessary to avoid redundant effort in the bug triaging and resolution process~\cite{32Bettenburg:ICSM08,Li2019,3Bettenburg:FSE08,Mao:ASE17}. In a user study which we detail later in this paper (Sec.~\ref{sec:user_study}), we investigated the effort required for experienced programmers to identify duplicate video-based bug reports and found that participants reported a range of difficulties for the task (\eg a single change of one step can result in two very similar looking videos showing entirely different bugs), and spent about $20$ seconds of comprehension effort on average per video viewed.
If this effort is extrapolated to the large influx of bug reports that could be collected on a daily basis \cite{Rakha:TSE'18,Rakha2018,32Bettenburg:ICSM08,Chaparro2016a}, it illustrates the potential for the excessive effort associated with video-based duplicate bug detection. This is further supported by the plans of a large company that supports crowd-sourced bug reporting (name omitted for anonymity), 
which we contacted as part of eliciting the design goals for this research, who stated that they anticipate increasing developer effort in managing video-based reports and that they are planning to build a feature in their framework to support this process.

To aid developers in determining whether video-based bug reports depict the same bug, this paper introduces \tango, a novel approach that analyzes both visual and textual information present in mobile screen-recordings using tailored computer vision (CV) and text retrieval (TR) techniques, with the goal of generating a list of candidate videos (from an issue tracker) similar to a target video-based report. In practice, \tango is triggered upon the submission of a new video-based report to an issue tracker. A developer would then use \tango to retrieve the video-based reports that are most similar (\eg top-5) to the incoming report for inspection. If duplicate videos are found in the ranked results, the new bug report can be marked as a duplicate in the issue tracker. Otherwise, the developer can continue to inspect the ranked results until she has enough confidence that the newly reported bug was not reported before (\ie it is not a duplicate).

\tango operates \textit{purely} upon the graphical information in videos in order to offer flexibility and practicality. These videos may show the \textit{unexpected behavior} of a mobile app (\ie a crash or other misbehavior) and the \textit{steps to reproduce} such behavior. Two videos are considered to be \textit{duplicates} if they show the same unexpected behavior (\aka a bug) regardless of the steps to reproduce the bug.
Given the nature of screen-recordings, video-based bug reports are likely to depict unexpected behavior towards the end of the video.
\tango attempts to account for this by leveraging the temporal nature of video frames and weighting the importance of frames towards the end of videos more heavily than other segments.

We conducted two empirical studies to measure: (i) the \textit{effectiveness} of different configurations of \tango by examining the benefit of combining visual and textual information from videos, as opposed to using only a single information source; and~(ii) \tangos ability to \textit{save developer effort} in identifying duplicate video-based bug reports. To carry out these studies, we collected a set of 180 video-bug reports from six popular apps used in prior research~\cite{Bernal-Cardenas:ICSE'20,Chaparro:FSE'19,Moran:FSE15,Moran:ICST16}, and defined 4,860 duplicate detection tasks that resemble those that developers currently face for textual bug reports -- wherein a corpus of potential duplicates must be ranked according to their similarity to an incoming bug report. 

The results of these studies illustrate that \tangos most effective configuration, which selectively combines visual and textual information, achieves 79.8\% mRR and 73.2\% mAP, an average rank of 1.7, a {\sc Hit}@1 of 67.7\%, and a {\sc Hit}@2 of 83\%. This means that \tango is able to suggest correct duplicate reports in the top-2 of the ranked candidates for 83\% of duplicate detection tasks.
The results of the user study we conducted with experienced programmers demonstrate that
on a subset of the tasks, \tango can reduce the time they spend in finding duplicate video-based bug reports by $\approx65\%$.

In summary, the main contributions of this paper are:
\begin{enumerate}
    \item{\tango, a duplicate detection approach for video-based bug reports of mobile apps which is able to accurately suggest duplicate reports;}
    \item{The results of a comprehensive empirical evaluation that measures the \textit{effectiveness} of \tango in terms of suggesting candidate duplicate reports;}
    \item{The results of a user study with experienced programmers that illustrates \tango's practical applicability by measuring its potential for \textit{saving developer effort}; and}
    \item{A benchmark (included in our online appendix \cite{appendix}) that enables (i) future research on video-based duplicate detection, bug replication, and mobile app testing, and (ii) the replicability of this work. The benchmark contains 180 video-based bug reports with duplicates, source code, trained models,  duplicate detection tasks, \tango's output, and detailed evaluation results.}
\end{enumerate}

\begin{figure*}[t]
	
	\centering
	\includegraphics[width=0.98\textwidth]{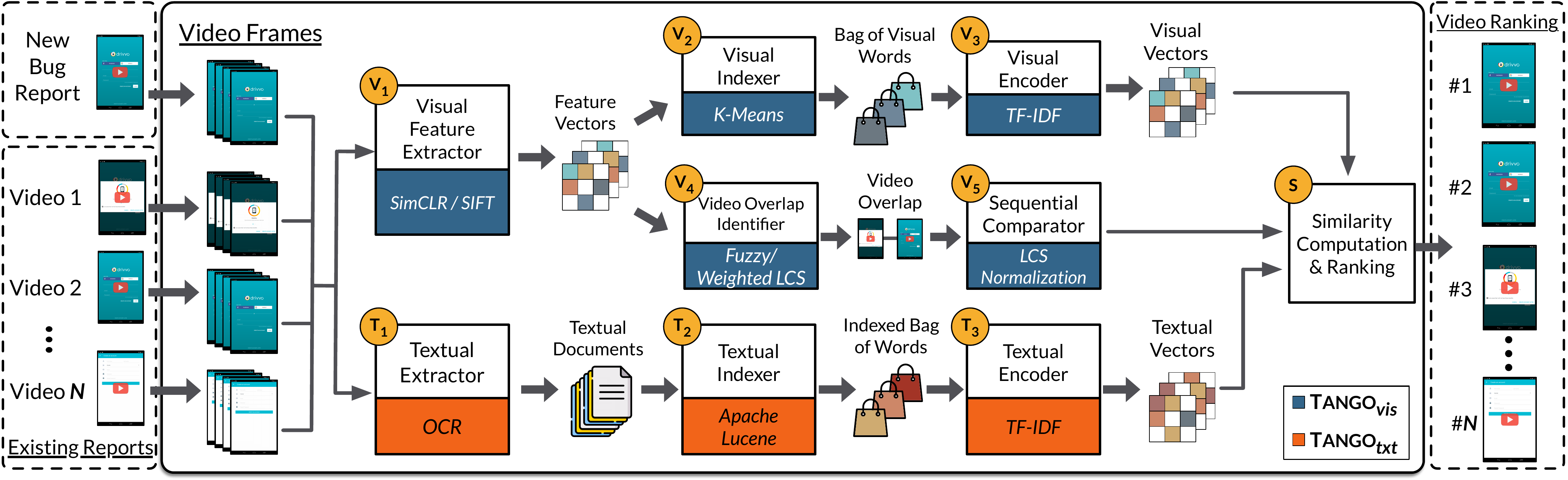}
	\caption{\normalsize{The \tango approach for detecting duplicate video-based bug reports.}}
	\vspace{-0.55cm}
	\label{fig:approach}
\end{figure*}

\section{The Tango Approach}\label{sec:approach}
\tango (\tangolong) is an automated approach based on CV and TR techniques, which leverages visual and textual information to detect duplicate video-based bug reports.

\subsection{\tango Overview}

\tango models duplicate bug report detection as an information retrieval problem. Given a new video-based bug report, \tango computes a similarity score between the new video and videos previously submitted by app users in a bug tracking system. The new video represents the query and the set of existing videos represent the corpus. \tango sorts the corpus of videos in decreasing order by similarity score and returns a ranked list of candidate videos. In the list, those videos which are more likely to show the same bug as the new video are ranked higher than those that show a different bug.

\tango has two major components, which we refer to as \tangov and \tangot (Fig.~\ref{fig:approach}), that compute video similarity scores independently. \tangov computes the \textit{visual similarity} and \tangot computes the \textit{textual similarity} between videos. The resulting similarity scores are linearly combined to obtain a final score that indicates the likelihood of two videos being duplicates. In designing \tangov, we incorporated support for two methods of computing visual similarity --- one of which is sensitive to the \textit{sequential order} of visual data, and the other one that is not --- and we evaluate the effectiveness of these two techniques in experiments described in Sec.~\ref{sec:eval}-\ref{sec:results}.

The first step in \tango's processing pipeline (Fig.~\ref{fig:approach}) is to decompose the query video, and videos from the existing corpus, into their constituent frames using a given sampling rate (\ie 1 and 5 frames per second - fps). Then, the \tangov and \tangot components of the approach are executed in parallel. The \textit{un-ordered} \tangov pipeline is shown at the top of Fig.~\ref{fig:approach}, comprising steps \circled{\textbf{\scriptsize V$_\textnormal{\textbf{1}}$}}-\circled{\textbf{\scriptsize V$_\textnormal{\textbf{3}}$}}; the \textit{ordered} \tangov pipeline is illustrated in the middle of Fig.~\ref{fig:approach}, comprising steps \circled{\textbf{\scriptsize V$_\textnormal{\textbf{1}}$}}, \circled{\textbf{\scriptsize V$_\textnormal{\textbf{4}}$}}, and \circled{\textbf{\scriptsize V$_\textnormal{\textbf{5}}$}}; and finally, the \tangot pipeline is illustrated at the bottom of Fig.~\ref{fig:approach} through steps \circled{\textbf{\scriptsize T$_\textnormal{\textbf{1}}$}}-\circled{\textbf{\scriptsize T$_\textnormal{\textbf{3}}$}}. Any of these three pipelines can be used to compute the video ranking independently or in combination (\ie combining the two \tangov together, one \tangov pipeline with \tangot, which we call \tangoc, or all three -- see Sec.~\ref{sec:configurations}). Next, we discuss these three pipelines in detail.

\subsection{T{\small ANGO}$_{vis}$: Measuring Unordered Visual Video Similarity}

The \textit{unordered} version of \tangov computes the visual similarity ($S_{vis}$) of video-based bug reports by extracting visual features from video frames and converting these features into a vector-based representation for a video using a Bag-of-Visual-Words (BoVW) approach \cite{Sivic:CCV'03,Jiang:IVR'07}. This process is depicted in the top of Fig. \ref{fig:approach}. The visual features are extracted by the \textit{visual feature extractor} model (\circled{\textbf{\scriptsize V$_\textnormal{\textbf{1}}$}} in Fig.~\ref{fig:approach}). Then, the \textit{visual indexer} \circled{\textbf{\scriptsize V$_\textnormal{\textbf{2}}$}} assigns to each frame feature vector a visual word from a visual codebook and produces a BoVW for a video. The \textit{visual encoder} \circled{\textbf{\scriptsize V$_\textnormal{\textbf{5}}$}}, based on the video BoVW, encodes the videos using a TF-IDF representation that can be used for similarity computation.

\subsubsection{Visual Feature Extraction}

 The \textit{visual feature extractor}~\circled{\textbf{\scriptsize V$_\textnormal{\textbf{1}}$}} can either use the SIFT~\cite{Lowe:JCV'04} algorithm to extract features, or SimCLR~\cite{Chen:SimCLR'20}, a recently proposed Deep Learning (DL) model capable of learning visual representations in an unsupervised, contrastive manner. \tango's implementation of SimCLR is adapted to extract visual features from app videos. 

The first method by which \tango can extract visual features is using the Scale-Invariant Feature Transform (SIFT)~\cite{Lowe:JCV'04} algorithm. SIFT is a state-of-the-art model for extracting local features from images that are invariant to scale and orientation. These features can be matched across images for detecting similar objects. This matching ability makes SIFT promising for generating features that can help locate duplicate images (in our case, duplicate video frames) by aggregating the extracted features. \tango's implementation of SIFT does not resize images and uses the top-10 features that are the most invariant to changes and are based on the local contrast of neighboring pixels, with higher contrast usually meaning more invariant. This is done to reduce the number of SIFT features, which could reach at least three orders of magnitude for a single frame, and make the \textit{visual indexing} \circled{\textbf{\scriptsize V$_\textnormal{\textbf{2}}$}} (through $k$-Means -- see Sec. \ref{sec:visual_indexing}) computationally feasible.

The other technique that \tango can use to extract features is SimCLR. In essence, the goal of this technique is to generate robust visual features that cluster similar images together while maximizing the distance between dissimilar images in an abstract feature space. This is accomplished by (i) generating sets of image pairs (containing one original image and one augmented image) and applying a variety of random augmentations (\ie image cropping, horizontal flipping, color jittering, and gray-scaling); (ii) encoding this set of image pairs using a base encoder, typically a variation of a convolutional neural network (CNN); and (iii) training a multi-layer-perceptron (MLP) to produce feature vectors that increase the cosine similarity between each pair of image variants and decrease the cosine similarity between negative examples, where negative examples for a given image pair are represented as all other images not in that pair, for a given training batch. %
 \tango's implementation of SimCLR employs the ResNet50~\cite{He:CVPR'16} CNN architecture as the base encoder, where this architecture has been shown to be effective~\cite{Chen:SimCLR'20}.
 
To ensure that \tango's \textit{visual feature extractor} is tailored to the domain of mobile app screenshots, we trained this component on the entire RICO dataset~\cite{Deka:UIST'17}, which contains over 66k Android screenshots from over 9k of the most popular apps on Google Play.  %
Our implementation of  SimCLR was trained using a batch size of $1,792$ and $100$ epochs, the same hyperparameters (\eg learning rate, weight decay, \etc) recommended by Chen \etal \cite{Chen:SimCLR'20} in the original SimCLR paper, and resized images to 224$\times$224 to ensure consistency with our base ResNet50 architecture. The training process was carried out on an Ubuntu 20.04 server with three NVIDIA T4 Tesla 16GB GPUs.
The output of the \textit{feature extractor} for SimCLR is a feature vector (of size 64) for each frame of a given video.

\subsubsection{Visual Indexing}
\label{sec:visual_indexing}

While the SimCLR or SIFT feature vectors generated by \tango's \textit{visual feature extractor} \circled{\textbf{\scriptsize V$_\textnormal{\textbf{1}}$}} could be used to directly compute the similarity between video frames, recent work has suggested that a BoVW approach combined with a TF-IDF similarity measure is more adept to the task of video retrieval~\cite{Kordopatis-Zilos:TM'19}. Therefore, to transform the SimCLR or SIFT feature vectors into a BoVW representation, \tango uses a \textit{visual indexing process} \circled{\textbf{\scriptsize V$_\textnormal{\textbf{2}}$}}.
This process produces an artifact known as a Codebook that maps SimCLR or SIFT feature vectors to ``visual words'' --- which are discrete representations of a given image, and have been shown to be suitable for image and video recognition tasks~\cite{Kordopatis-Zilos:TM'19}. %
The Codebook derives these visual words by clustering feature vectors and computing the centroids of these clusters, wherein each centroid corresponds to a different visual word. %

The Codebook makes use of the $k$-Means clustering algorithm, where the $k$ represents the diversity of the visual words, and thus can affect the representative power of this indexing process. \tango's implementation of this process is configurable to 1k, 5k, or 10k for the $k$ number of clusters (\ie the number of visual words - VW). 1k VW and 10k VW were selected as recommended by Kordopatis-Zilos \etal \cite{Kordopatis-Zilos:TM'19} and we included 5k VW as a ``middle ground'' to better understand how the number of visual words impacts \tango's performance. A Codebook is generated only once for a given $k$, however, it must be trained before it can be applied to convert an input feature vector to its corresponding visual word(s). Once trained, a Codebook can then be used to map visual words from frame feature vectors without any further modification. Thus, we trained \tango's six Codebooks, three for SIFT and three for SimCLR, using features extracted from $15,000$ randomly selected images from the RICO dataset~\cite{Deka:UIST'17}. We did not use the entire RICO dataset due to computational constraints of the $k$-means algorithm. %

After the feature vector for a video frame is passed through the \textit{visual indexing} process, it is mapped to its BoVW representation by a trained Codebook. To do this, the Codebook selects the closest centroid to each visual feature vector, based on Euclidean distance. For SIFT, this process may generate more than one feature vector for a single frame, due to the presence of multiple SIFT feature descriptors. In this case, \tango assigns multiple visual words to each frame. For SimCLR, \tango assigns one visual word to each video frame, as SimCLR generates only one vector per frame.

\subsubsection{Visual Encoding} After the video is represented as a BoVW, the \textit{visual encoder} \circled{\textbf{\scriptsize V$_\textnormal{\textbf{3}}$}} computes the final vector representation of the video through a TF-IDF-based approach~\cite{Salton:TFIDF86}. The term frequency (TF) is computed as the number of visual words occurrences in the BoVW representation of the video, and the inverse document frequency (IDF) is computed  as the number of occurrences of a given visual word in the BoVW representations built from the RICO dataset. Since RICO does not provide videos but individual app screenshots, we consider each RICO image as one document. We opted to use RICO to compute our IDF representation for two reasons: (i) to combat the potentially small number of frames present in a given video recording, and (ii) to bolster the generalizability of our similarity measure across a diverse set of apps.

\subsubsection{Similarity Computation}

Given two videos, \tangov encodes them into their BoVW representations, and each video is represented as one visual TF-IDF vector. These vectors are compared 
using cosine similarity, which is taken as the \textit{visual similarity} \circled{\textbf{\scriptsize S}} of the videos ($S_{vis}=S_{BoVW}$).

\subsection{T{\small ANGO}$_{vis}$: Measuring Ordered Visual Video Similarity}

The \textit{ordered} version of \tangov considers the sequence of video frames when comparing two videos and is capable of giving more weight to common frames nearer the end of the videos, as this is likely where buggy behavior manifests. 
To accomplish this, the \textit{feature vector extractor} \circled{\textbf{\scriptsize V$_\textnormal{\textbf{1}}$}} is used to derive descriptive vectors from each video frame using either SimCLR or SIFT. \tango determines how much the two videos overlap using an adapted longest common substring (LCS) algorithm \circled{\textbf{\scriptsize V$_\textnormal{\textbf{4}}$}}. Finally, during the \textit{sequential comparison process} \circled{\textbf{\scriptsize V$_\textnormal{\textbf{5}}$}}, \tango calculates the similarity score by normalizing the computed LCS score.
\subsubsection{Video Overlap Identification} 

In order to account for the sequential ordering of videos, \tango employs two different versions of the longest common substring (LCS) algorithm. The first version, which we call fuzzy-LCS (f-LCS), modifies the comparison operator of the LCS algorithm to perform fuzzy matching instead of exact matching between frames in two videos. This fuzzy matching is done differently for SimCLR and SIFT-derived features. For SimCLR, given that each frame is associated with only a single visual word, the standard BoVW vector would be too sparse for a meaningful comparison. Therefore, we compare the feature vectors that SimCLR extracts from the two frames \textit{directly} using cosine similarity. For SIFT, we utilize the BoVW vectors derived by the \textit{visual encoder} \circled{\textbf{\scriptsize V$_\textnormal{\textbf{3}}$}}, but at a per-frame level.

The second LCS version, which we call weighted-LCS~(w-LCS), uses the same fuzzy matching that f-LCS performs. However, the similarity produced in this matching is then weighted depending on where the two frames from each video appeared. Frames that appear later in the video are weighted more heavily, since that is where the buggy behavior is typically occurring in a video-based bug report, and thus should be given more weight for duplicate detection. The exact weighting scheme used is $\frac{i}{m} \times \frac{j}{m}$, where $i$ is the $ith$ frame of video A, $m$ is the \# of frames in video A, $j$ is the $jth$ frame of video B, and $n$ is the \# of frames in video B.

\subsubsection{Sequential Comparison}

In order to incorporate the LCS overlap measurements into \tango's overall video similarity calculation, the overlap scores must be normalized between zero and one ($[0,1]$). To accomplish this, we consider the case where two videos overlap perfectly to be the upper bound of the possible LCS score between two videos, and use this to perform the normalization. For f-LCS, this is done by simply dividing by the \# of frames in the smaller video since the $max$ possible overlap that could occur is when the smaller video is a subsection in the bigger video, calculated as $overlap/min$ where $overlap$ denotes the amount the two videos share in terms of their frames and $min$ denotes the \# of frames in the smaller of the two videos. For w-LCS, if the videos are different lengths, we align the end of the shorter video to the \textit{end} of the longer video and consider this the upper bound on the LCS score, which is normalized as follows:

\begin{equation}
\label{eq:wlcs}
   S_{w-LCS} = \frac{overlap}{\sum_{i=min}^{1}\frac{i}{min}\times\frac{max - i}{max}}
\end{equation}

\noindent where $S_{w-LCS}$ is the normalized similarity value produced by w-LCS, $overlap$ and $min$ are similar to the f-LCS calculation and $max$ denotes the \# of frames in the longer of the two videos. The denominator in Eq. \ref{eq:wlcs} calculates the maximum possible overlap that can occur if the videos were exact matches, summing across the similarity score of each frame pair. Our online appendix contains the detailed f/w-LCS algorithms with examples \cite{appendix}.

\subsubsection{Similarity Computation} f-LCS and w-LCS output the \textit{visual similarity} \circled{\textbf{\scriptsize S}} score $S_{f-LCS}$ and $S_{w-LCS}$, respectively. This can be combined with $S_{BoVW}$ to obtain an aggregate visual similarity score:  $S_{vis} = (S_{BoVW} + S_{f-LCS})/2$ or $S_{vis} = (S_{BoVW} + S_{w-LCS})/2$. We denote these \tangov variations as \bflcs and \bwlcs, respectively.

\subsection{Determining the Textual Similarity between Videos}
\label{sec:approach_textual}

In order to determine the textual similarity between video-based bug reports, \tango leverages the textual information from labels, titles, messages, \etc found in the app GUI components and screens depicted in the videos.

\tangot adopts a standard text retrieval approach based on Lucene \cite{Hatcher2004} and Optical Character Recognition (OCR)~\cite{tesseract-ocr,pytesseract} to compute the textual similarity ($S_{txt}$) between video-based bug reports. First, a textual document is built from each video in the issue tracker (\circled{\textbf{\scriptsize T$_\textnormal{\textbf{1}}$}} in Fig. \ref{fig:approach}) using OCR to extract text from the video frames. The textual documents are pre-processed using standard techniques to improve similarity computation, namely tokenization, lemmatization, and removal of punctuation, numbers, special characters, and one- and two-character words. The pre-processed documents are indexed for fast similarity computation \circled{\textbf{\scriptsize T$_\textnormal{\textbf{2}}$}}. Each document is then represented as a vector using TF-IDF and the index~\cite{Salton:TFIDF86} \circled{\textbf{\scriptsize T$_\textnormal{\textbf{3}}$}}. 

In order to build the textual documents from the videos, \tangot applies OCR on the video frames through the Tesseract engine \cite{tesseract-ocr,pytesseract} in the \textit{textual extractor} \circled{\textbf{\scriptsize T$_\textnormal{\textbf{1}}$}}. We experiment with three strategies to compose the textual documents using the extracted frame text. The first strategy (all-text) concatenates all the text extracted from the frames. The second strategy (unique-frames) concatenates all the text extracted from unique video frames, determined by applying exact text matching (before text pre-processing). The third strategy (unique-words) concatenates the unique words in the frames (after pre-processing).

\subsubsection{Similarity Computation} \tango computes the \textit{textual similarity} ($S_{txt}$) in \circled{\textbf{\scriptsize S}} using Lucene's scoring function~\cite{lucene-tfidfsimilarity} based on cosine similarity and document length normalization.

\subsection{Combining Visual and Textual Similarities}
\label{sec:approach_combination}
\tango combines both the visual ($S_{vis}$) and textual ($S_{txt}$) similarity scores produced by \tangov and \tangot, respectively (\circled{\textbf{\scriptsize S}} in Fig. \ref{fig:approach}). \tango uses a linear combination approach to produce an aggregate similarity value:
\begin{equation}\label{eq:pythagoras}
S_{comb} = (1 - w) \times S_{vis} + w \times S_{txt}
\end{equation}
where $w$ is a weight for $S_{vis}$ and $S_{txt}$, and takes a value between zero (0) and one (1). Smaller $w$ values weight $S_{vis}$ more heavily, and larger values weight $S_{txt}$ more heavily. We denote this approach as \tangoc.

Based on the combined similarity, \tango generates a ranked list of the video-based bug reports found in the issue tracker. This list is then inspected by the developer to determine if a new video reports a previously reported bug.

\section{\tango's Empirical Evaluation Design}\label{sec:eval}

We empirically evaluated \tango with two goals in mind: (i) determining how effective \tango is at detecting duplicate video-based bug reports, when considering different configurations of components and parameters, and (ii) estimating the effort that \tango can save developers during duplicate video bug detection. Based on these goals, we defined the following research questions (RQs):

\begin{enumerate}[label=\textbf{RQ$_\arabic*$:}, ref=\textbf{RQ$_\arabic*$}, wide, labelindent=5pt]\setlength{\itemsep}{0.2em}
    \item \label{rq:individual_performance}{\textit{How effective is \tango when using either visual or textual information alone to retrieve duplicate videos?}}
    \item \label{rq:lcs_performance}{\textit{What is the impact of combining frame sequence and visual information on \tango's detection performance?}}
    \item \label{rq:combination_performance}{\textit{How effective is \tango when combining both visual and textual information for detecting duplicate videos?}}
    \item \label{rq:time}{\textit{How much effort can \tango{} save developers in finding duplicate video-based bug reports?}}
\end{enumerate}

To answer our \textbf{RQs}, we first collected video-based bug reports for six Android apps (Sec.~\ref{sec:data_collection}), and based on them, defined a set of duplicate detection tasks (Sec.~\ref{sec:scenarios_and_tasks}). We instantiated different configurations of \tango by combining its components and parameters (Sec.~\ref{sec:configurations}), and executed these configurations on the defined tasks (Sec.~\ref{sec:evaluation_methodology}). Based on standard metrics, applied on the video rankings that \tango produces, we measured \tango's effectiveness (Sec.~\ref{sec:evaluation_methodology}). We answer \ref{rq:individual_performance}, \ref{rq:lcs_performance}, and \ref{rq:combination_performance} based on the collected measurements. To answer \ref{rq:time}  (Sec.~\ref{sec:user_study}), we conducted a user study where we measured the time humans take to find duplicates for a subset of the defined tasks, and estimated the time \tango can save for developers. We present and discuss the evaluation results in Sec.~\ref{sec:results}.

\subsection{Data Collection}
\label{sec:data_collection}

We collected video-based bug reports for six open-source Android apps, namely AntennaPod (APOD)~\cite{antennapod}, Time Tracker (TIME)~\cite{time-tracker}, Token (TOK)~\cite{token}, GNUCash (GNU)~\cite{gnucash}, GrowTracker (GROW)~\cite{growtracker}, and Droid Weight (DROID)~\cite{droidweight}. We selected these apps because they have been used in previous studies \cite{Bernal-Cardenas:ICSE'20,Chaparro:FSE'19,Moran:FSE15,Moran:ICST16}, support different app categories (finance, productivity, \etc), and provide features that involve a variety of GUI interactions (taps, long taps, swipes, \etc). Additionally, none of these apps are included as part of the RICO dataset used to train \tango's SimCLR model and Codebooks, preventing the possibility of data snooping. Since video-based bug reports are not readily available in these apps' issue trackers, we designed and carried out a systematic procedure for collecting them. 

In total, we collected 180 videos that display 60 distinct bugs -- 10 bugs for each app and three videos per bug (\ie three duplicate videos per bug). From the 60 bugs, five bugs (one bug per app except for DROID) are reported in the apps' issue trackers. These five bugs were selected
because they were the only ones in the issue trackers that we were able to reproduce based on the provided bug descriptions. During the reproduction process, we discovered five \textit{additional} new bugs in the apps not reported in the issue trackers (one bug each for APOD, GNU, and TOK, and two bugs for TIME) for a total of 10 confirmed real bugs.

The remaining 50 bugs were introduced in the apps through mutation by executing MutAPK~\cite{Escobar-Velasquez:ASE'19}, a mutation testing tool that injects bugs (\ie mutants) into Android APK binaries via a set of 35 mutation operators that were derived from a large-scale empirical study on \textit{real} Android application faults. Given the empirically-derived nature of these operators, they were shown to accurately simulate real-world Android faults~\cite{Escobar-Velasquez:ASE'19}. We applied MutAPK to the APKs of all six apps. Then, from the mutant list produced by the tool, we randomly selected 7 to 10 bugs for each app, and ensured that they could be reproduced and manifested in the GUI. To diversify the bug pool, we selected the bugs from multiple mutant operators and ensured that they affected multiple app features/screens.

When selecting the 60 bugs, we ensured they manifest graphically and were reproducible by manually replicating them on a specific Android emulator configuration (virtual Nexus 5X with Android 7.0 configured via Android Studio). For all the bugs, we screen-recorded the bug and the reproduction scenario. We also generated a textual bug report (for bugs that did not have one) containing the description of the unexpected and expected app behavior and the steps to reproduce the bug.

To generate the remaining 120 video-based bug reports, we asked two professional software engineers and eight computer science (CS) Ph.D. students to replicate and record the bugs (using the same Android emulator), based only on the textual description of the unexpected and expected app behavior.
The participants have between 2 and 10 years of programming experience (median of 6 years). 

All the textual bug reports given to the study participants contained \textit{only} a brief description of the observed and expected app behavior, with \textit{no specific reproduction steps}.
We opted to perform the collection in this manner to ensure the robustness of our evaluation dataset by maximizing the diversity of video-based reproduction steps, and simulating a real-world scenario where users are aware of the observed (incorrect) and expected app behavior, and must produce the reproduction steps themselves.

We randomly assigned the bugs to the participants in such a way that each bug was reproduced and recorded by two participants, and no participant recorded the same bug twice. Before reproducing the bugs, the participants spent five minutes exploring the apps to become familiar with their functionality. Since some of the participants could not reproduce the bugs after multiple attempts (mainly due to bug misunderstandings) and some of the videos were incorrectly recorded (due to mistakes), we reassigned these bugs among the other participants, who were able to reproduce and record them successfully.

Our bug dataset consists of 35 crashes and 25 non-crashes, and include a total of 470 steps (397 taps, 12 long taps, 14 swipes, among other types), with an average of 7.8 steps per video. The average video length is $\approx28$ seconds.

\subsection{Duplicate Detection Tasks}
\label{sec:scenarios_and_tasks}

For each app, we defined a set of tasks that mimic a realistic scenario for duplicate detection.  
Each duplicate detection task is composed of a new video (\ie the new bug report, \aka the query) and a set of existing videos (\ie existing bug reports in the issue tracker, \aka the corpus). In practice, a developer would determine if the new video is a duplicate by inspecting the corpus of videos in the order given by \tango (or any other approach). For our task setup, the corpus contains both duplicate and non-duplicate videos. There are two different types of duplicate videos that exist in the corpus: (i) those videos that are a duplicate of the query (the \textit{Same Bug} group), and (ii) those videos which are duplicates of each other, but are not a duplicate of the query (the \textit{Different Bug} group). This second type of duplicate video is represented by bug reports marked as duplicates in the issue tracker and their corresponding master reports \cite{Rakha:TSE'18,Sun2011,Chaparro2016a}.
Each non-duplicate video reports a distinct bug.

We constructed the duplicate detection tasks on a \textit{per app} basis, using the 30 video reports collected for each app (\ie three video reports for each of the 10 bugs, for a total of 30 video reports per app). We first divided all the 30 videos for an app into three groups, each group containing 10 videos (one for each bug) created by one or more participants. Then, we randomly selected a video from one bug as the query and took the other two videos that reproduce the same bug as the \textit{Same Bug} duplicate group (\ie the ground truth). Then, we selected one of the remaining nine bugs and added its three videos to the \textit{Different Bug} duplicate group. Finally, we selected one video from the remaining eight bugs, and used these as the corpus' \textit{Non-Duplicate} group. This resulted in a total of 14 distinct bug reports per task (two in the \textit{Same Bug} group, three in the \textit{Different Bug} group, eight in the \textit{Non-Duplicate} group, and the query video). After creating tasks based on all the combinations of query and corpus, we generated a total of $810$ duplicate detection tasks per app or $4,860$ aggregating across all apps.

We designed the duplicate detection setting described above to mimic a scenario likely to be encountered in crowd-sourced app testing, where duplicates of the query, other duplicates not associated with the query, and other videos reporting unique bugs, exist in a shared corpus for a given app. While there are different potential task settings, we opted not to vary this experimental variable in order to make for a feasible analysis that allowed us to explore more thoroughly the different \tango configurations.

\subsection{\tango Configurations}
\label{sec:configurations}

We designed \tangov and \tangot to have different configurations. \tangov's configurations are based on different visual feature extractors (SIFT or SimCLR), video sampling rates (1 and 5 fps), \# of visual words (1k, 5k, and 10k VW), and approaches to compute video similarity (BoVW, f-LCS, w-LCS, \bflcs, and \bwlcs). \tangot's configurations are based on the same sampling rates (1 and 5 fps) and the approaches to extract the text from the videos (all-text, unique-frames, and unique-words). \tangoc combines \tangov and \tangot as described in Sec.~\ref{sec:approach_combination}.

\subsection{\tango's Execution and Effectiveness Measurement}
\label{sec:evaluation_methodology}

We executed each \tango configuration on the $4,860$ duplicate detection tasks and measured its effectiveness using standard metrics used in prior text-based duplicate bug detection research~\cite{Rakha:TSE'18,Sun2011,Chaparro2016a}. For each task, we compare the ranked list of videos produced by \tango and the expected duplicate videos from the ground truth.

We measured the \textit{rank} of the first duplicate video found in the ranked list, which serves as a proxy for how many videos the developer has to watch in order to find a duplicate video. A smaller \textit{rank} means higher duplicate detection effectiveness. Based on the \textit{rank}, we computed the \textit{reciprocal rank} metric: $1/rank$. We also computed the \textit{average precision} of \tango, which is the average of the precision values achieved at all the cutting points k of the ranked list (\ie precision@k). Precision@k is the proportion of the top-k returned videos that are duplicates according to the ground truth. We also computed \textit{HIT@k} (\aka Recall Rate@k \cite{Rakha:TSE'18,Sun2011,Chaparro2016a}), which is the proportion of tasks that are successful for the cut point k of the ranked list. A task is successful if at least one duplicate video is found in the top-k results returned by \tango. We report HIT@k for cut points k = 1-2 in this paper, and 1-10 in our online appendix~\cite{appendix}.

Additionally, we computed the average of these metrics over sets of duplicate detection tasks: mean reciprocal rank (mRR), mean average precision (mAP), and mean rank ($\mu$ rank or $\mu$Rk) per app and across all apps. Higher mRR, mAP, and HIT@k values indicate higher duplicate detection effectiveness. These metrics measure the overall performance of a duplicate detector.

We focused on comparing mRR values to decide if one \tango configuration is more effective than another, as we consider that it
better reflects the usage scenario of \tango. In practice, the developer would likely stop inspecting the suggested duplicates (given by \tango) when she finds the first correct duplicate. This scenario is captured by mRR, through the \textit{rank} metric, which considers only the first correct duplicate video as opposed to the entire set of duplicate videos to the query (as mAP does).

\subsection{Investigating \tango's Effort Saving Capabilities}
\label{sec:user_study}

We conducted a user study in order to estimate the effort that developers would spend while manually finding video-based duplicates. This effort is then compared to the effort measurements of the best \tango configuration, based on $\mu$~\textit{rank} and \textit{HIT@k}. This study and the data collection procedure were conducted remotely due to COVID-19 constraints.

\subsubsection{Participants and Tasks}
One professional software engineer and four CS Ph.D. students from the data collection procedure described in Sec.~\ref{sec:data_collection} participated in this study. The study focused on APOD, the app that all the participants had in common from the data collection. We randomly selected 20 duplicate detection tasks, covering all 10 APOD bugs. 

\subsubsection{Methodology}

Each of the 20 tasks was completed by two participants. Each participant completed four tasks, each task's query video reporting a unique bug. The assignment of the tasks to the participants was done randomly. For each task, the participants had to watch the new video (the query) and then find the videos in the corpus that showed the same bug of the new video (\ie find the duplicate videos). All the videos were anonymized so that the duplicate videos were unknown to the participants. To do this, we named each video with a number that represents the video order and the suffix ``vid'' (\eg ``2\_vid.mp4'').

The corpus videos were given in random order and the participants could watch them in any order. To make the bug of the new video clearer to the participants, we provided them with the description of the unexpected and expected app behavior, taken from the textual bug reports that we generated for the bugs. We consider the randomization of the videos as a reasonable baseline given that other baselines (\eg video-based duplicate detectors) do not currently exist and the video-based bug reports in our dataset do not have timestamps (which can be used to give a different order to the videos). This is a threat to validity that we discuss in Sec. \ref{sec:limits}.

\subsubsection{Collected Measurements}

Through a survey, we asked each participant to provide the following information for each task: (i) the name of the first video they deemed a duplicate of the query, (ii) the time they spent to find this video, (iii) the number of videos they had to watch until finding the first duplicate (including the duplicate), (iv) the names of other videos they deemed duplicates, and (v) the time they spent to find these additional duplicates. We instructed the participants to perform the tasks without any interruptions in order to minimize inaccuracies in the time measurements.

\subsubsection{Comparing \tango and Manual Duplicate Detection}

The collected measurements from the participants were compared against the effectiveness obtained by executing the best \tango configuration on the 20 tasks, in terms of $\mu$~\textit{rank} and \textit{HIT@k}. We compared the avg. number of videos the participants watched to find one duplicate against the avg. number of videos they would have watched had they used~\tango.

\section{\tango's Evaluation Results}
\label{sec:results}

\subsection{RQ1: Using Only Visual or Textual Information}

\begin{table}[t]
\centering
\small
\caption{Effectiveness for the best  \tango configurations that use either visual (SimCLR/SIFT) or textual (OCR\&IR) information.}
\label{tab:all_best_indiv}
\begin{tabular}{C{7.5mm}|c|ccc|cc}
\toprule
\textbf{App}             & \textbf{Config.} & \textbf{mRR} & \textbf{mAP} & \textbf{\footnotesize{$\mu$Rk}} & \textbf{\footnotesize{HIT@1}} & \textbf{\footnotesize{HIT@2}} \\ \midrule
\multirow{3}{*}{\footnotesize{APOD}}    & \footnotesize{SIFT}           & 64.6\%       & 51.1\%       & 3.0             & 47.7\%       & 71.7\%       \\
                         				& \footnotesize{SimCLR}         & 80.0\%       & 66.8\%       & 1.7             & \textbf{68.1\%}       & 82.6\%       \\
                         				& \footnotesize{OCR\&IR}        & \textbf{80.8\%}       & \textbf{75.3\%}       & \textbf{1.5}             & 65.7\%       & \textbf{88.6\%}      \\ \hline
\multirow{3}{*}{\footnotesize{DROID}}   & \footnotesize{SIFT}           & 66.3\%       & 55.0\%       & 2.5             & 49.1\%       & 69.5\%       \\
                        				& \footnotesize{SimCLR}         & 64.6\%       & 59.2\%       & 2.6             & 49.5\%       & 61.7\%       \\
                        				& \footnotesize{OCR\&IR}        & \textbf{67.9\%}       & \textbf{64.7\%}       & \textbf{2.3}             & \textbf{52.0\%}       & \textbf{69.8\%}       \\ \hline
\multirow{3}{*}{\footnotesize{GNU}}     & \footnotesize{SIFT}           & 66.1\%       & 57.2\%       & 2.2             & 47.4\%       & 68.4\%       \\
                         				& \footnotesize{SimCLR}         & 81.8\%       & 75.1\%       & 1.6             & 70.1\%       & 85.3\%       \\
                         				& \footnotesize{OCR\&IR}        & \textbf{84.5\%}       & \textbf{82.3\%}       & \textbf{1.4}             & \textbf{72.2\%}       & \textbf{92.0\%}       \\ \hline
\multirow{3}{*}{\footnotesize{GROW}}    & \footnotesize{SIFT}           & 56.0\%       & 49.9\%       & 3.0             & 36.5\%       & 54.3\%       \\
                         				& \footnotesize{SimCLR}         & 72.7\%       & 68.8\%       & 2.0             & 57.4\%       & 75.6\%       \\
                         				& \footnotesize{OCR\&IR}        & \textbf{76.8\%}      & \textbf{69.0\%}       & \textbf{1.9}             & \textbf{63.6\%}       & \textbf{80.1\%}       \\ \hline
\multirow{3}{*}{\footnotesize{TIME}}    & \footnotesize{SIFT}           & 49.2\%       & 40.7\%       & 3.3             & 26.7\%       & 46.4\%       \\
                         				& \footnotesize{SimCLR}         & \textbf{74.8\%}       & \textbf{67.6\%}       & \textbf{2.3}             & \textbf{63.7\%}       & \textbf{75.9\%}       \\
                         				& \footnotesize{OCR\&IR}        & 47.4\%       & 37.7\%       & 4.0             & 28.3\%       & 44.4\%       \\ \hline
\multirow{3}{*}{\footnotesize{TOK}}     & \footnotesize{SIFT}           & 39.0\%       & 32.1\%       & 4.4             & 17.0\%       & 33.7\%       \\
                         				& \footnotesize{SimCLR}         & \textbf{77.7\%}       & \textbf{69.3\%}       & \textbf{1.6}             & \textbf{60.6\%}       & \textbf{86.7\%}       \\
                         				& \footnotesize{OCR\&IR}        & 61.3\%       & 53.3\%       & 2.6             & 42.6\%       & 60.7\%       \\ \midrule
\multirow{3}{*}{\footnotesize{\hspace{-0.1cm}\textbf{Overall}}} & \footnotesize{SIFT}           & 56.9\%       & 47.7\%       & 3.1             & 37.4\%       & 57.3\%       \\
                                        & \footnotesize{SimCLR}         & \textbf{75.3\%}       & \textbf{67.8\%}       & \textbf{1.9}             & \textbf{61.6\%}       & \textbf{78.0\%}       \\
                         				& \footnotesize{OCR\&IR}        & 69.8\%       & 63.7\%       & 2.3             & 54.1\%       & 72.6\%       \\ \bottomrule
\end{tabular}
\end{table}

We analyzed the performance of \tango 
when using only visual or textual information exclusively. In this section, we present the results for \tango's best performing configurations. However, complete results can be found in our online appendix~\cite{appendix}.
Table \ref{tab:all_best_indiv} shows the results for \tangov and \tangot when using SimCLR, SIFT, as the \textit{visual feature extractor}, and OCR as the \textit{textual extractor}. For simplicity, we use SimCLR, SIFT, and OCR\&IR to refer to SimCLR-based \tangov,  SIFT-based \tangov, and \tangot, respectively. The best results for each metric are illustrated in bold on a per app basis.
The results provided in Table \ref{tab:all_best_indiv} are those for the best parameters of the SimCLR, SIFT, and OCR\&IR feature extractors, which are (BoVW, 5 fps, 1k VW), (w-LCS, 1 fps, 10k VW), and (all-text, 5 fps), respectively.

Table \ref{tab:all_best_indiv} shows that \tangov is more effective when using SimCLR rather than SIFT across all the apps, achieving an overall mRR, mAP, avg. rank, HIT@1, and HIT@2 of 75.3\%, 67.8\%, 1.9, 61.6\%, and 78\%, respectively. SimCLR is also superior to OCR\&IR overall, whereas SIFT performs least effectively of the three approaches.
When analyzing the results per app, we observe that SimCLR is outperformed by OCR\&IR (by 0.7\% - 4\% difference in mRR) for APOD, DROID, GNU and GROW; with OCR\&IR being the most effective for these apps. SimCLR outperforms the other two approaches for TIME and TOK by more than 16\% difference in mRR. The differences explain the overall performance of SimCLR and OCR\&IR. SimCLR is more consistent in its performance compared to OCR\&IR and SIFT. Across apps, the mRR standard deviation of SimCLR is 6.2\%, which is lower than that for SIFT and OCR\&IR: 11.1\% and 13.9\%, respectively. The trend is similar for mAP and avg. rank. 

Since the least consistent approach across apps is \tangot in terms of effectiveness, we investigated the root causes for its lower performance on TIME and TOK.  After manually watching a subset of the videos for these apps, we found that their textual content was quite similar across bugs.
Based on this, we hypothesized that the amount of vocabulary shared between duplicate videos (from the same bugs) and non-duplicate videos (across different bugs) affected the discriminatory power of Lucene-based \tangot (see Sec.~\ref{sec:approach_textual}).

To verify this hypothesis, we measured the shared vocabulary of duplicate and non-duplicate video pairs, similarly to Chaparro \etal's analysis of textual bug reports \cite{Chaparro2016a}. We formed unique pairs of duplicate and non-duplicate videos from all the videos collected for all six apps. For each app, we formed 30 duplicate and 405 non-duplicate pairs, and we measured the avg. amount of shared vocabulary of all pairs, using the vocabulary agreement metric used by Chaparro \etal~\cite{Chaparro2016a}. Table \ref{tab:vocab_agreement} shows the vocabulary agreement of duplicate ($V_d$) and non-duplicate pairs ($V_{nd}$) as well as the mRR and mAP values of  \tangot  for each app. The table reveals that the vocabulary agreement of duplicates and non-duplicates is very similar for TIME and TOK, and dissimilar for the other apps. The absolute difference between these measurements (\ie $|V_d - V_{nd}|$) for TIME and TOK is 0.3\% and 8.6\%, while for the other apps it is above 16\%. We found 0.94 / 0.91 Pearson correlation~\cite{Freedman:07} between these differences and the mRR/mAP values. 

The results indicate that, for TIME and TOK, the similar vocabulary between duplicate and non-duplicate videos negatively affects the discriminatory power of \tangot, which suggests that for some apps, using only textual information may be sub-optimal for duplicate detection.

\begin{table}[t]
\centering
\small
\caption{Vocabulary agreement \& effectiveness for the best \tangot.} 
\label{tab:vocab_agreement}
\begin{tabular}{c|ccc|cc}
\toprule
\multirow{2}{*}{\textbf{App}} & \multicolumn{3}{c|}{\textbf{Vocabulary agreement}} & \multirow{2}{*}{\textbf{mRR}} & \multirow{2}{*}{\textbf{mAP}} \\ \cline{2-4}
        & \textbf{$V_d$} & \textbf{$V_{nd}$} & \textbf{$|V_d - V_{nd}|$} &        &        \\ 
\midrule
APOD    & 70.8\%         & 37.9\%            & 32.9\%                    & 80.8\% & 75.3\% \\
DROID   & 73.9\%         & 57.0\%            & 16.9\%                    & 67.9\% & 64.7\% \\
GNU     & 82.2\%         & 58.6\%            & 23.6\%                    & 84.5\% & 82.3\% \\
GROW    & 67.0\%         & 41.7\%            & 25.4\%                    & 76.8\% & 69.0\% \\
TIME    & 86.0\%         & 86.3\%            & 0.3\%                     & 47.4\% & 37.7\% \\
TOK     & 69.6\%         & 61.0\%            & 8.6\%                     & 61.3\% & 53.3\% \\ 
\midrule
\textbf{Overall} & 74.2\%         & 56.7\%            & 17.5\%                    & 69.8\% & 63.7\% \\ 
\bottomrule
\end{tabular}
\end{table}

\begin{tcolorbox}[enhanced,skin=enhancedmiddle,borderline={1mm}{0mm}{MidnightBlue}]
\textbf{Answer for \ref{rq:individual_performance}}: SimCLR performs the best overall with an mRR and HIT@1 of 75.3\% and 61.6\%, respectively. For 4 of 6 apps, OCR\&IR outperforms SimCLR by a significant margin. However, due to issues with vocabulary overlap, it performs worse overall. SIFT is the worst-performing technique across all the apps.
\end{tcolorbox}

\subsection{RQ2: Combining Visual and Frame Sequence Information}

To answer \ref{rq:lcs_performance}, we compared the effectiveness of the best configuration of \tango when using visual information alone (SimCLR, BoVW, 5fps, 1k VW) and when combining visual \& frame sequence information~(\ie \bflcs and~\bwlcs).

The results are shown in Table \ref{tab:all_best_lcs}. Overall, using \tango with BoVW alone is more effective than combining the approaches; \tango based on BoVW achieves 75.3\%, 67.8\%, 1.9, 61.6\%, and 78\% mRR, mAP, avg. rank, HIT@1, and HIT@2, respectively. Using BoVW and w-LCS combined is the least effective approach. BoVW alone and \bflcs are comparable in performance. However, BoVW is more consistent in its performance across apps: 6.2\% mRR std. deviation vs. 6.6\% and 9.2\% for  \bflcs and \bwlcs.

The per-app results reveal that \bwlcs consistently is the least effective approach for all apps except for GROW, for which \bwlcs performs best. After watching the videos for GROW, we found unnecessary steps in the beginning/middle of the duplicate videos, which led to their endings being weighted more heavily by w-LCS, where steps were similar. In contrast, BoVW and \bflcs give a lower weight to these cases thus reducing the overall video similarity. 

The lower performance of  \bflcs and \bwlcs, compared to BoVW, is partially explained by the fact that f-LCS and w-LCS are more restrictive by definition. Since they find the longest common sub-strings of frames between videos, small variations (\eg extra steps) in the reproduction steps of the bugs may lead to drastic changes in similarity measurement for these approaches. Also, these approaches only find one common substring (\ie the longest one), which may not be highly discriminative for duplicate detection. In the future, we plan to explore additional approaches for aligning the frames, for example, by using an approach based on longest common sub-sequence algorithms \cite{Gusfield:LCS'97} that can help align multiple portions between videos. Another potential reason for these results may lie in the manner that \tango combines visual and sequential similarity scores -- weighting both equally. In future work, we plan to explore additional combination techniques.

\begin{table}[t]
\centering
\small
\caption{Effectiveness for the best \tangov configuration using either visual information (BoVW) or a combination of visual and frame sequence information (\bflcs and \bwlcs).}
\vspace{-0.5em}
\label{tab:all_best_lcs}
\begin{tabular}{C{7.5mm}|c|ccc|cc}
\toprule
\textbf{App}             & \textbf{Config.} & \textbf{mRR} & \textbf{mAP} & \textbf{\footnotesize{$\mu$Rk}} & \textbf{\footnotesize{HIT@1}} & \textbf{\footnotesize{HIT@2}} \\ \midrule
\multirow{3}{*}{\footnotesize{APOD}}    & \footnotesize{\bflcs}      & 79.3\%       & \textbf{67.8\%}      & \textbf{1.7}             & 66.2\%       & 82.3\%           \\
                         				& \footnotesize{\bwlcs}      & 77.2\%       & 65.5\%      & 1.9             & 64.2\%       & 80.1\%      \\
                         				& \footnotesize{BoVW}        & \textbf{80.0\%}       & 66.8\%      & \textbf{1.7}             & \textbf{68.1\%}       & \textbf{82.6\%}    \\ \hline
\multirow{3}{*}{\footnotesize{DROID}}   & \footnotesize{\bflcs}      & \textbf{64.8\%}       & \textbf{60.7\%}      & \textbf{2.6}             & \textbf{50.2\%}       & 61.6\%       \\
                         				& \footnotesize{\bwlcs}      & 63.7\%       & 54.8\%      & 2.7             & 48.9\%       & 62.3\%        \\
                         				& \footnotesize{BoVW}        & 64.6\%       & 59.2\%      & \textbf{2.6}             & 49.5\%       & \textbf{61.7\%}     \\ \hline
\multirow{3}{*}{\footnotesize{GNU}}     & \footnotesize{\bflcs}      & \textbf{83.3\%}       & \textbf{75.6\%}      & \textbf{1.6}             & \textbf{73.2\%}       & \textbf{85.6\%}       \\
                         				& \footnotesize{\bwlcs}      & 77.3\%       & 65.7\%      & 1.8             & 62.3\%       & 83.6\%          \\
                         				& \footnotesize{BoVW}        & 81.8\%       & 75.1\%      & \textbf{1.6}             & 70.1\%       & 85.3\%        \\ \hline
\multirow{3}{*}{\footnotesize{GROW}}    & \footnotesize{\bflcs}      & 76.0\%       & 70.2\%      & 2.0             & 64.2\%       & 75.2\%        \\
                         				& \footnotesize{\bwlcs}      & \textbf{81.3\%}       & \textbf{75.0\%}      & \textbf{1.7}             & \textbf{70.9\%}       & \textbf{82.8\%}         \\
                         				& \footnotesize{BoVW}        & 72.7\%       & 68.8\%      & 2.0             & 57.4\%       & 75.6\%       \\ \hline
\multirow{3}{*}{\footnotesize{TIME}}    & \footnotesize{\bflcs}      & 70.4\%       & 63.4\%      & \textbf{2.3}             & 54.4\%       & 74.3\%        \\
                         				& \footnotesize{\bwlcs}      & 63.8\%       & 58.5\%      & 2.8             & 48.0\%       & 64.9\%        \\
                         				& \footnotesize{BoVW}        & \textbf{74.8\%}       & \textbf{67.6\%}      & \textbf{2.3}             & \textbf{63.7\%}       & \textbf{75.9\%}        \\ \hline
\multirow{3}{*}{\footnotesize{TOK}}     & \footnotesize{\bflcs}      & 73.4\%       & 65.6\%      & 1.7             & 54.0\%       & 82.5\%       \\
                         				& \footnotesize{\bwlcs}      & 59.2\%       & 53.7\%      & 2.6             & 37.9\%       & 60.0\%       \\
                         				& \footnotesize{BoVW}        & \textbf{77.7\%}       & \textbf{69.3\%}      & \textbf{1.6}             & \textbf{60.6\%}       & \textbf{86.7\%}         \\ \midrule
\multirow{3}{*}{\footnotesize{\hspace{-0.1cm}\textbf{Overall}}} & \footnotesize{\bflcs}      & 74.5\%       & 67.2\%      & 2.0             & 60.4\%       & 76.9\%        \\
                         				& \footnotesize{\bwlcs}      & 70.4\%       & 62.2\%      & 2.2             & 55.4\%       & 72.3\%        \\
                         				& \footnotesize{BoVW}        & \textbf{75.3\%}       & \textbf{67.8\%}      & \textbf{1.9}             & \textbf{61.6\%}       & \textbf{78.0\%}      \\ \bottomrule
\end{tabular}
\end{table}

\begin{tcolorbox}[enhanced,skin=enhancedmiddle,borderline={1mm}{0mm}{MidnightBlue}]
\textbf{Answer for \ref{rq:lcs_performance}}: Combining  ordered visual information (via f-LCS and w-LCS) with the orderless BoVW improves the results for four of the six apps. However, across all apps, BoVW performs more consistently.
\end{tcolorbox}

\subsection{RQ3: Combining Visual and Textual Information}
\label{sec:results_combined}

\begin{table}[t]
\centering
\small
\caption{Effectiveness of the best \tangoc, \tangov, and \tangot.}
\vspace{-0.5em}
\label{tab:combined_best}
\begin{tabular}{C{7.5mm}|c|ccC{4.5mm}|C{8mm}C{8mm}}
\toprule
\textbf{App}             & \textbf{Config.} & \textbf{mRR} & \textbf{mAP} & \textbf{\footnotesize{$\mu$Rk}} & \textbf{\footnotesize{HIT@1}} & \textbf{\footnotesize{HIT@2}} \\ \midrule
\multirow{3}{*}{\footnotesize{APOD}}  & \footnotesize{\tangoc} 	   & \textbf{84.4\%}       & \textbf{75.8\%}       & \textbf{1.4}             & \textbf{73.1\%}       & 87.9\%         \\ 
                       				  & \footnotesize{\tangov}       & 80.0\%       & 66.8\%       & 1.7             & 68.1\%       & 82.6\%          \\ 
                      			      & \footnotesize{\tangot}       & 80.8\%       & 75.3\%       & 1.5             & 65.7\%       & \textbf{88.6\%}          \\ \hline
\multirow{3}{*}{\footnotesize{DROID}} & \footnotesize{\tangoc} 	   & \textbf{70.6\%}       & \textbf{66.7\%}       & \textbf{2.2}             & \textbf{55.9\%}       & \textbf{71.0\%}         \\ 
                       				  & \footnotesize{\tangov}       & 64.6\%       & 59.2\%       & 2.6             & 49.5\%       & 61.7\%        \\ 
                       				  & \footnotesize{\tangot}       & 67.9\%       & 64.7\%       & 2.3             & 52.0\%       & 69.8\%          \\ \hline
\multirow{3}{*}{\footnotesize{GNU}}   & \footnotesize{\tangoc} 	   & \textbf{89.5\%}       & \textbf{84.7\%}       & \textbf{1.3}             & \textbf{81.6\%}       & \textbf{94.2\%}         \\ 
                       			      & \footnotesize{\tangov}       & 81.8\%       & 75.1\%       & 1.6             & 70.1\%       & 85.3\%          \\ 
                       				  & \footnotesize{\tangot}       & 84.5\%       & 82.3\%       & 1.4             & 72.2\%       & 92.0\%           \\ \hline
\multirow{3}{*}{\footnotesize{GROW}}  & \footnotesize{\tangoc} 	   & \textbf{81.7\%}       & \textbf{75.4\%}       & \textbf{1.7}             & \textbf{71.4\%}       & \textbf{82.5\%}          \\ 
                       				  & \footnotesize{\tangov}       & 72.7\%       & 68.8\%       & 2.0             & 57.4\%       & 75.6\%          \\ 
                       			 	  & \footnotesize{\tangot}       & 76.8\%       & 69.0\%       & 1.9             & 63.6\%       & 80.1\%          \\ \hline
\multirow{3}{*}{\footnotesize{TIME}}  & \footnotesize{\tangoc} 	   & 59.6\%       & 51.7\%       & 2.8             & 40.2\%       & 58.8\%          \\ 
                      				  & \footnotesize{\tangov}       & \textbf{74.8\%}       & \textbf{67.6\%}       & \textbf{2.3}             & \textbf{63.7\%}       & \textbf{75.9\%}        \\ 
                      				  & \footnotesize{\tangot}       & 47.4\%       & 37.7\%       & 4.0             & 28.3\%       & 44.4\%       \\ \hline
\multirow{3}{*}{\footnotesize{TOK}}   & \footnotesize{\tangoc} 	   & 69.8\%       & 60.8\%       & 2.0             & 50.9\%       & 76.9\%            \\ 
                     				  & \footnotesize{\tangov}       & \textbf{77.7\%}       & \textbf{69.3\%}       & \textbf{1.6}             & \textbf{60.6\%}       & \textbf{86.7\%}          \\ 
                     				  & \footnotesize{\tangot}       & 61.3\%       & 53.3\%       & 2.6             & 42.6\%       & 60.7\%          \\ \midrule
\multirow{3}{*}{\footnotesize{\hspace{-0.1cm}\textbf{Overall}}}   & \footnotesize{\tangoc}   & \textbf{75.9\%}       & \textbf{69.2\%}       & \textbf{1.9}             & \textbf{62.2\%}       & \textbf{78.5\%}           \\ 
                      				  & \footnotesize{\tangov}       & 75.3\%       & 67.8\%       & \textbf{1.9}             & 61.6\%       & 78.0\%            \\ 
                     				  & \footnotesize{\tangot}       & 69.8\%       & 63.7\%       & 2.3             & 54.1\%       & 72.6\%           \\
                       \bottomrule
\end{tabular}
\end{table}

We investigated \tango's effectiveness when combining visual and textual information. We selected the best configurations of \tangov (SimCLR, BoVW,  5  fps,  1k  VW) and \tangot (all-text, 5 fps) from \ref{rq:individual_performance} based on their mRR score and measured its performance overall and per app. We provide the results for the best weight we obtained for \tango's \textit{similarity computation and ranking} which was $w=0.2$, \ie a weight of 0.8 and 0.2 on \tangov and \tangot, respectively. These weights were found by evaluating different $w$ values from zero ($0$) to one ($1$) in increments of $0.1$ and selecting the one leading to the highest overall mRR score. Complete results can be found in our online appendix \cite{appendix}.

Table \ref{tab:combined_best} shows that the overall effectiveness achieved by \tangoc is higher than that achieved by \tangot and \tangov. \tangoc achieves 75.9\%, 69.2\%, 1.9, 62.2\%, and 78.5\% mRR, mAP, avg. rank, HIT@1, and HIT@2, on average. The avg. improvement margin of \tangoc is substantially higher for \tangot (6.2\%/5.5\% mRR/mAP) than for \tangov  (0.7\%/1.4\% mRR/mAP). 

Our analysis of the per-app results explains these differences. Table \ref{tab:combined_best} reveals that combining visual and textual information substantially increases the performance over just using one of the information types alone, except for the TIME and TOK apps. This is because \tangot's effectiveness is substantially lower for these apps, compared to the visual version (see Table \ref{tab:all_best_indiv}), due to the aforementioned vocabulary agreement. Thus, incorporating the textual information significantly harms the performance of \tangoc.

\subsubsection{A Better Combination of Visual and Textual Information}
  
The results indicate that combining visual and textual information is beneficial for most of our studied apps but harmful for a subset (TIME and TOK). This is because the textual information used alone, for TIME and TOK, leads to low performance. The analysis we made for \tangot in \ref{rq:individual_performance}, revealed that the reason for the low performance of \tangot lies in the similar amount of vocabulary overlap between duplicate and non-duplicate videos. Fortunately, based on this amount of vocabulary, we can predict the performance of \tangot for new video-based bug reports as follows \cite{Chaparro2016a}. In practice, the issue tracker will contain reports marked as duplicates (reporting the same bugs) from previous submissions of bug reports as well as non-duplicates (reporting unique bugs). This information can be used to compute the vocabulary agreement between duplicates and non-duplicates, which can be used to predict how well \tangot would perform for new reports.

Based on this, we defined a new approach for \tango, which is based on the vocabulary agreement metric from \cite{Chaparro2016a} applied on existing duplicate and non-duplicate reports. This approach dictates that if the difference of vocabulary agreement between existing duplicates and non-duplicates is greater than a certain threshold, then \tango should combine visual and textual information. Otherwise, \tango should only use the visual information because it is likely that the combination would not be better than using the visual information alone.

From the vocabulary agreement measurements shown in Table \ref{tab:vocab_agreement}, we infer a proper threshold from the new \tango approach. This threshold may be taken as one value from the interval 8.6\% - 16.9\% (exclusive) because those are the limits that separate the apps for which \tangot obtains low (TIME and TOK) and high performance (APOD, DROID, GNU, and GROW). For practical reasons, we select the threshold to be the middle value: $8.6 + (16.9 - 8.6)/2 = 12.8\%$. In future work, we plan to further evaluate this threshold on other apps.

We implemented this approach for \tango, using 0.2 as weight, and measured its effectiveness. This approach resulted in a mRR, mAP, avg. rank, HIT@1 and HIT@2 of 79.8\%, 73.2\%, 1.7, 67.7\%, and 83\%, respectively. The approach leads to a substantial improvement (\ie 3.9\% / 4.1\% higher mRR / mAP) over \tangoc shown in Table~\ref{tab:combined_best}.

The results mean that the best version of \tango is able to suggest correct duplicate video-based bug reports in the first or second position of the returned candidate list for 83\% of the duplicate detection tasks.

\begin{tcolorbox}[enhanced,skin=enhancedmiddle,borderline={1mm}{0mm}{MidnightBlue}]
\textbf{Answer for \ref{rq:combination_performance}}: Combining visual and textual information significantly improves results for 4 of 6 apps. However, due to the vocabulary agreement issue, across all apps, this approach is similar in effectiveness to using visual information alone. Accounting for this vocabulary overlap issue through a selective combination of visual and textual information via a threshold, \tango achieves the highest effectiveness: an mRR, mAP, avg. rank, HIT@1, and HIT@2 of 79.8\%, 73.2\%, 1.7, 67.7\%, and 83\%, respectively.
\end{tcolorbox}

\subsection{RQ4: Time Saved Discovering Duplicates}
\label{sec:results_user_study}

As expected, the participants were successful in finding the duplicate videos for all 20 tasks. In only one task, one participant incorrectly flagged a video as duplicate because it was highly similar to the query. Participants found the first duplicate video in 96.4 seconds and watched 4.3 videos on avg. across all tasks to find it. Participants also found all the duplicates in 263.8 seconds on avg. by watching the entire corpus of videos. This means they spent 20.3 seconds in watching one video on average.

We compared these results with the measurements taken from \tango's best version (\ie selective \tango) on the tasks the participants completed. \tango achieved a 1.5 avg. rank, which means that, by using \tango, they would only have to watch  one or two videos on avg. to find the first duplicate. This would have resulted in $(4.3-1.5) / 4.3 =65.1\%$ of the time saved. In other words, instead of spending $20.3 \times 4 = 81.2$ seconds (on avg.) finding a duplicate for a given task, the participants could have spent $20.3 \times 1.5 = 30.5$ seconds. These results indicate the potential of \tango to help developers save time when finding duplicates.

\begin{tcolorbox}[enhanced,skin=enhancedmiddle,borderline={1mm}{0mm}{MidnightBlue}]
\textbf{Answer for \ref{rq:time}}: On average, \tango's best-performing configuration can save 65.1\% of the time participants spend finding duplicate videos.
\end{tcolorbox}

\vspace{-0.15cm}
\section{\tango Limitations \& Threats to Validity}\label{sec:limits}

\textbf{Limitations.} \tango has three main limitations that motivate future work. 
 
The first one stems from the finding that textual information may not be beneficial for some apps. The best \tango version implements an approach for detecting this situation, based on a threshold for the difference in vocabulary overlap between duplicate and non-duplicate videos, which is used for selectively combining visual or textual information.
This threshold is based on the collected data and may not generalize to other apps. Second, 
the visual TF-IDF representation for the videos is based on the mobile app images from the RICO dataset, rather than on the videos found in the tasks' corpus, as it is typically done in text retrieval. Additionally, we considered single images as documents rather than groups of frames that  make up a video. These decisions were made to improve the generalization of \tango's visual features and to support projects that have limited training data. Third, differences in themes and languages across video-based bug reports for an application could have an impact in the performance of \tango. We believe that different themes (\ie dark vs. light modes) will not significantly impact \tango since the SimCLR model is trained to account for such color differences by performing color jittering and gray-scaling augmentations. However, additional experiments are needed to validate this conjecture. For different languages, \tango currently assumes the text in an application to be English when performing OCR and textual similarity. Therefore, its detection effectiveness where the bug reports display different languages (\eg English vs. French) could be negatively impacted. We will investigate this aspect in our future work.

\textbf{Internal \& Construct Validity.} 
Most of the mobile app bugs in our dataset were introduced by MutAPK \cite{Escobar-Velasquez:ASE'19}, and hence potentially may not resemble real bugs. However, MutAPK's mutation operators were empirically derived from a large-scale study of real Android faults, and prior research lends credence of the ability of mutants to resemble real faults~\cite{Andrews:ICSE05}. We intentionally selected generated mutants from a range of operators to increase the diversity of our set of bugs and mitigate potential biases. Another potential threat is related to using real bugs from issue trackers that cannot be reproduced or that do not manifest graphically. We mitigated this threat by using a small, carefully vetted subset of real bugs that were analyzed by multiple authors before being used in our dataset. 
We did not observe major differences in the results between mutants and real bugs.

Another threat to validity is that our approach to construct the duplicate detection tasks does not take into account bug report timestamps, which would be typical in a realistic scenario~\cite{Rakha:TSE'18}, and timestamps could be used as a baseline ordering of videos for comparing against the ranking given by \tango. The lack of timestamps stems from the fact that we were not able to collect the video-based bug reports from existing mobile projects. We mitigated this threat in our user study by randomizing the ordering of the corpus videos given to the participants. We consider this as a reasonable baseline for evaluating our approach considering that, to the best of our knowledge, (1) no existing datasets, with timestamps, are available for conducting research on video-based duplicate detection, and (2) no existing duplicate detectors work exclusively on video-based bug reports, as \tango does.

\textbf{External Validity.} We selected a diverse set of apps that have different functionality, screens, and visual designs, in an attempt to mitigate threats to external validity. Additionally, our selection of bugs also attempted to select diverse bug types (crashes and non-crashes), and the duplicate videos were recorded by different participants. As previously discussed, there is the potential that \tango's different parameters \& thresholds may not generalize to video data from other apps.

\section{Related Work}
\label{sec:rel-work}

Our research is related to work in near duplicate video retrieval, analysis of graphical software artifacts, and duplicate detection of textual bug reports.

\textbf{Near Duplicate Video Retrieval.} Extensive research has been done outside SE in near-duplicate video retrieval, which is the identification of similar videos to a query video (\eg exact copies \cite{Douze:TM'10,Kraaij:11,Jiang:16,Hao:TM'17} or similar  events \cite{Chen:ICDM'06,Jiang:IVR'07,Revaud:CVPR'13,Kordopatis-Zilos:TM'19}). 

The closest work to ours is by Kordopatis-Zilos \etal~\cite{Kordopatis-Zilos:TM'19}, who addressed the problem of retrieving videos of incidents (\eg accidents). 
In their work, they explored using handcrafted-based \cite{Lowe:JCV'04,Bay:ECCV'06,Wu:MM'07,Zhao:PAMI'07,Jing:19,Huang:CVPR'97} (\eg SURF or HSV histograms) and DL-based \cite{Chechik:JMLR'10,Kordopatis-Zilos:ICCVW'17,Lecun:CNN'98,Kordopatis-Zilos:17,Tran:ICCV'15,Carreira:17} (\eg CNNs) visual feature extraction techniques and ways of combining the extracted visual features \cite{Jegou:CVPR'10,Revaud:CVPR'13,Sivic:CCV'03,Jiang:IVR'07,Cai:MM'11,Kordopatis-Zilos:17} (\eg VLAD).
While we do make use of the best performing model (CNN+BoVW) from this work~\cite{Kordopatis-Zilos:TM'19}, we did not use the proposed handcrafted approaches, as these were designed for scenes about real-world incidents, rather than for mobile bug reporting. We also further modified and extended this approach given our different domain, through the combination of visual and textual information modalities, and adjustments to the CCN+BoVW model, including the layer configuration and training objective.

\textbf{Analysis of Graphical Software Artifacts.} The analysis of graphical software artifacts
to support software engineering tasks has been common in recent years. Such tasks include mobile app testing \cite{Jones:2014,Moran:ICST16,Hu:FSE18,Bernal-Cardenas:ICSE'20}, developer/user behavior modelling \cite{Caetano:02,Bao:ICSE15,Frisson:CHI16}, GUI reverse engineering and code generation \cite{Dixon:11,Nguyen:ASE15,Beltramelli:EICS18,Chen:ICSE18,Moran:TSE18,Chen:FSE20}, analysis of programming videos \cite{MacLeod:ICPC'15,Lasecki:ACM'15,Yadid:2016,Ponzanelliz:TSE'19,Alahmadi:EMSE20,Zhao:ICSE19}, and GUI understanding and verification \cite{Chang:ACM'11,Zhao:ICSE20}. None of these works deal with finding duplicate video-based bug reports, which is our focus.

\textbf{Detection of Duplicate Textual Bug Reports.}
Many research projects have focused on detecting duplicate textual bug reports
\cite{Bettenburg2008,Borg2014,Chaparro2016a,Chaparro2019,He2020,Hindle2016,Hindle2018,Klein2014,Lazar2014,Lerch2013a,Liu2013,Nguyen2012,Panichella2019,Rakha2018,Rakha:TSE'18,Rodrigues2020,Runeson2007,Sun2010,Sun2011,Sureka2010,Thung2014,Tian2012,Wang2008,Wang2019,Wang2020,Zhou2012a}. Similar to \tango, most of the proposed techniques return a ranked list of duplicate candidates \cite{Kang2017,Chaparro2016a}.  The work most closely related to \tango is by Wang \etal \cite{Wang2019}, who leveraged attached mobile app images to better detect duplicate textual reports. Visual features are extracted from the images (\eg representative colors), using computer vision, which are combined with textual features extracted from the text to obtain an aggregate similarity score. While this is similar to our work, \tango is intended to be applied to videos rather than single images and focuses on video-based bug reports alone, without any extra information such as bug descriptions.

\section{Conclusion and Future Work}\label{sec:conclusion}

This paper presented \tango, an approach that combines visual and textual information to help developers find duplicate video-based bug reports. Our empirical evaluation, conducted on $4,680$ duplicate detection tasks created from 180 video-based bug reports from six mobile apps, illustrates that \tango is able to effectively identify duplicate reports and save developer effort. Specifically, \tango correctly suggests duplicate video-based bug reports within the top-2 candidate videos for 83\% of the tasks, and saves  65.1\% of the time that humans spend finding duplicate videos.

Our future work will focus on addressing \tango's limitations and extending \tango's evaluation. Specifically, we plan to (1) explore additional ways to address the vocabulary overlap problem, (2) investigate the resilience of \tango to different app characteristics such as the use of different themes, languages, and screen sizes, (3) extend \tango for detecting duplicate bug reports that contain multimedia information (text, images, and videos), (4) evaluate \tango using data from additional apps, and (5) assess the usefulness of \tango in industrial settings.

\section{Data Availability}\label{sec:data}
Our online appendix~\cite{appendix} includes the collected video-based bug reports with duplicates, \tango's source code, trained models, evaluation infrastructure, \tango's output, and detailed evaluation results.

\section*{Acknowledgements}
\label{sec:acknowledgement}
We thank Winson Ye for experimenting with different CNN models, the study participants for their time and useful feedback, and the companies who responded to our initial internal survey. This research was supported in part by the NSF CCF-1955853 and CCF-1815186 grants. Any opinions, findings, and conclusions expressed herein are the authors’ and do not necessarily reflect those of the sponsors.

\IEEEpeerreviewmaketitle

\balance
\bibliographystyle{abbrv}
\bibliography{references.bib,references_dup_detection.bib}

\end{document}